\documentclass[11pt,a4paper,twoside,groupcitations]{article}
\usepackage[T1]{fontenc}
\usepackage[ansinew]{inputenc}
\usepackage[english]{babel}
\usepackage{amsfonts}
\usepackage{amsmath}
\usepackage{bm}
\usepackage{array}
\usepackage{amsthm}
\usepackage{amssymb}
\usepackage{graphicx}
\usepackage{subfigure}
\usepackage{braket}
\usepackage{eucal}
\usepackage{verbatim}
\usepackage[table]{xcolor}
\usepackage{caption}
\usepackage{cite}
\usepackage{textcomp}
\usepackage{multicol}
\usepackage{tikz}
\usetikzlibrary{positioning,arrows}
\usetikzlibrary{decorations.pathmorphing}
\usetikzlibrary{decorations.markings}
\usetikzlibrary{calc,decorations.markings}
\usetikzlibrary{arrows,shapes}
\usetikzlibrary{matrix,arrows}
\usepackage{pgfplots}
\usepackage{xparse}
\definecolor{jade}{HTML}{00A86B}

\newcommand{\ba}{\begin{eqnarray}}
\newcommand{\ea}{\end{eqnarray}}

\def\ben{\begin{equation}}
\def\een{\end{equation}}
\def\bea{\begin{eqnarray}}
\def\eea{\end{eqnarray}}
\def\be{\begin{equation}}
\def\ee{\end{equation}}

\def\nn{\nonumber}

\newcommand{\hoch}[1]{$\, ^{#1}$}

\def\ben{\begin{equation}}
\def\een{\end{equation}}
\def\bea{\begin{eqnarray}}
\def\eea{\end{eqnarray}}
\def \nn {\nonumber}

\def\ft#1#2{{\textstyle{\frac{\scriptstyle #1}{\scriptstyle #2} } }}
\def\fft#1#2{{\frac{#1}{#2}}}

\begin{document}


\begin{center}

{\Large {\bf Blandford-Znajek Process in Einsteinian Cubic Gravity}}

\vspace{40pt}
{\bf Jun Peng\hoch{1\dagger} and Xing-Hui Feng\hoch{2*}}

\vspace{10pt}

\hoch{1}{\it Faculty of Science, Beijing University of Technology,
Beijing 100124, China}

\hoch{2}{\it Center for Joint Quantum Studies and Department of Physics,
School of Science, Tianjin University, Tianjin 300350, China}

\vspace{40pt}

\underline{ABSTRACT}
\end{center}

In this paper, we investigate the Blandford-Znajek (BZ) process within the framework of Einsteinian cubic gravity (ECG). To analytically study the BZ process using the split monopole configuration, we construct a slowly rotating black hole in ECG up to cubic order in small spin, considering the leading order in small coupling constant of higher curvature terms. By deriving the magnetosphere solution around the black hole, we determine the BZ power up to the second relative order in spin. The BZ power is modified by the coupling constant compared to Kerr black hole. Although the general nature of the BZ process in ECG remains unchanged at the leading order in spin, the coupling constant introduces modification at the second relative order in spin. Therefore, we anticipate that it is feasible to discern general relativity from higher derivative gravities by examining the BZ power in rapidly rotating black holes.

\vfill {\footnotesize \hoch{\dagger} junpeng@bjut.edu.cn \ \ \hoch{*} xhfeng@tju.edu.cn}

\thispagestyle{empty}

\pagebreak




\section{Introduction}
In 1977 Blandford and Znajek (BZ) proposed a mechanism that efficiently extracts the rotational energy of black holes by magnetic field \cite{Blandford:1977ds}. The BZ process is realizable in an astrophysical environment, since the magnetic fields can be supported by accretion disks around stellar-mass and supermassive black holes (BH). Thus the BZ process is thought to be the most promising candidate for the power source of ultra high energy cosmic rays, such as gamma-ray bursts and relativistic jets from active galactic nuclei. In the BZ process, the ergosphere of a rotating BH is immersed in a poloidal magnetic field. Due to frame-dragging, the magnetic field lines twist in the toroidal direction, resulting in the emergence of a poloidal electric field.  In this way the rotational energy of a BH is transferred into the energy of the currents outside the BH.

This process has been extensively studied by analytical method \cite{Blandford:1977ds,Beskin:2000qe,McKinney:2004ka,Tanabe:2008wm,Tchekhovskoy:2009ba,Pan:2015haa,Pan:2015iaa,Grignani:2018ntq,Grignani:2019dqc,Armas:2020mio} and numerical stimulation \cite{Komissarov:2001,Komissarov:2004qu,Tchekhovskoy:2009ba,Tchekhovskoy:2011zx,Nathanail:2014aua} during the past four decades in general relativity (GR). The simplest analytical study of the BZ mechanism is based on the "split monopole" configuration \cite{Blandford:1977ds}. These studies allow for the computation of the fields perturbatively up to a particular order of the BH's spin, making them valid only for slowly rotating BH. For the investigation of BZ process in rapidly rotating BH, numerical simulations are usually necessary. Recently Armas et al. have extended the perturbative scheme to arbitrary orders of spin and any magnetic field configuration \cite{Armas:2020mio}, and the perturbative expansions were improved in \cite{Camilloni:2022kmx}.

Since the BZ process depends on the ergosphere of BH, studying the process and its observational signatures can provide insights into gravity in the strong-field regime. Specifically, investigating the BZ process in deformed Kerr geometry, resulting from modifications of GR or certain astrophysical environments, can help constrain the metric deformation parameters using BZ power. Some attempts were made in this direction, such as studying BZ process for parametrically deformed Kerr BHs \cite{Bambi:2012zg,Pei:2016kka,Konoplya:2021qll}, for Kerr-Sen BH in heterotic string theory \cite{Banerjee:2020ubc} and for slowly rotating BH in scalar Gauss-Bonnet gravity and dynamical Chern-Simons gravity \cite{Dong:2021yss}.

The analysis in \cite{Dong:2021yss} shows a degeneracy between BH spin and the coupling constants of modified terms at leading order in slow rotation approximation which is broken at higher orders. As a result, if we want to distinguish GR from other theories of gravity using BZ power, we must study the BZ process to high orders of spin. Roughly speaking, amplitude of rotation amplifies the differences between GR and modified gravities. Another evidence for this cognition can be found in recent paper \cite{Horowitz:2023xyl}. In this paper, we study the BZ process to the second relative order in spin for Einsteinian cubic gravity (ECG) which is a well-motivated high derivative gravity \cite{Bueno:2016xff}. This can be a further example that enhances the comprehension of how to learn about new physics from BH observations related to BZ mechanism.

This paper is organized as follows. In section 2 we give a brief introduction to BZ mechanism. In section 3 we construct slowly rotating BH in ECG. Then we obtain magnetosphere solution around the BH in section 4. In section 5 we analysis the BZ power in ECG. We conclude in section 6.

\section{Blandford-Znajek process}
In this section, we show the main ingredients of Blandford-Znajek process. In the split monopole configuration, the accretion disk is considered as a thin current sheet in the equatorial plane of the BH, and the magnetosphere around the BH is determined by force-free electrodynamics which implies \cite{Gralla:2014yja}
\be
F_{\mu\nu}J^\nu=0 \label{FFE}
\ee
where $F_{\mu\nu}$ is the field strength tensor, and $J^\nu=\nabla_\mu F^{\mu\nu}$ is the four-current.

In Boyer-Lindquist coordinates ($t,r,\theta,\phi$), a stationary and axisymmetric metric can be decomposed in the following form
\be
ds^2=g_{\mu\nu}dx^\mu dx^\nu=g^T_{AB}dx^Adx^B+g^P_{ab}dx^adx^b
\ee
$g^T_{AB}$ is referred to as the toroidal metric $(t, \phi)$, and $g^P_{ab}$ is referred to as the poloidal metric $(r, \theta)$. A stationary and axisymmetric electromagnetic field can always be represented by \cite{Gralla:2014yja}
\bea
&&F_{tr}=-F_{rt}=\Omega\partial_r\psi\nn\\
&&F_{t\theta}=-F_{\theta t}=\Omega\partial_\theta\psi\nn\\
&&F_{r\phi}=-F_{\phi r}=\partial_r\psi\nn\\
&&F_{\theta\phi}=-F_{\phi\theta}=\partial_\theta\psi\nn\\
&&F_{r\theta}=-F_{\theta r}=\frac{I}{2\pi}\sqrt{\frac{g^P}{-g^T}}
\eea
with all other components zero. Here $\psi, I, \Omega$ are functions of  $r, \theta$. The $t$ and $\phi$ components of the force-free condition \eqref{FFE} imply that $I, \Omega$ are functions of $\psi$. To summarize, the force-free electromagnetic field is characterized by three quantities: the magnetic flux $\psi(r,\theta)$ and the electric current $I(\psi)$ through a surface bounded by the loop of revolution at ($r,\theta$), and the angular velocity of magnetic field lines $\Omega(\psi)$ being dragged by the rotation of the BH. The $r$ and $\theta$ components of the force-free conditions \eqref{FFE} give the so-called stream equation
\be
\nabla_\mu(|\eta|^2\nabla^\mu\psi)+\Omega'(\eta\cdot dt)|\nabla\psi|^2-\frac{II'}{4\pi^2g^T}=0 \label{streameq}
\ee
where the prime denotes a derivative with respect to $\psi$, and $\eta\equiv d\phi-\Omega dt$.

To solve the above stream equation, we must impose reasonable boundary conditions for $\psi$. Here we follow a recent work by Armas et al. \cite{Armas:2020mio}, imposing the boundary conditions for $\psi$ as
\bea
\psi=0, &&\theta=0,\nn\\
\psi=\psi_0, &&\theta=\pi/2,\nn\\
\psi~\text{finite}, &&r=r_H,\nn\\
\psi~\text{finite}, &&r\rightarrow\infty,\label{bc}
\eea
where $\psi_0$ is the monopole charge, and $r_H$ is the horizon radius. These are natural boundary conditions for split monopole configuration. The solutions of $I, \Omega$ can be obtained by the so-called Znajek conditions \cite{Znajek(1977),Gralla:2014yja}
\bea
I=2\pi(\Omega-\Omega_H)\partial_\theta\psi\sqrt{\frac{g_{\phi\phi}}{g_{\theta\theta}}},&&r=r_H\nn\\
I=-2\pi\Omega\partial_\theta\psi\sin\theta,&&r\rightarrow\infty\label{znajek}
\eea
where $\Omega_H$ is the horizon angular velocity. The Znajek conditions are derived from the requirement that fields are finite in regular coordinates. Having obtained the solutions of field variables $\psi, I, \Omega$, the total electromagnetic energy flux extracted from the BH, also known as the BZ power, is given by \cite{Gralla:2014yja}
\be
P=-\int I\Omega d\psi=4\pi\int_0^{\pi/2}\left.\left[\Omega(\Omega_H-\Omega)(\partial_\theta\psi)^2\sqrt{\frac{g_{\phi\phi}}{g_{\theta\theta}}}\right]\right|_{r=r_H}d\theta\label{BZpower}
\ee


\section{Slowly rotating black hole in Einsteinian cubic gravity}
The action of Einsteinian cubic gravity includes a particular combination of cubic curvature terms \cite{Bueno:2016xff}. The ECG has two remarkable properties. First the vacuum static spherically symmetric solution is characterized by only one function in four dimension, which is determined by a two derivatives differential equation. Second, the linearized equations of motion on maximally symmetric backgrounds coincide with the linearized Einstein equations up to an overall factor. Thus it is free of massive spin-2 ghost for general higher curvature gravity. The action of ECG is
\be
{\cal S}=\frac{1}{16\pi G}\int dx^4 \sqrt{-g} (R+\lambda {\cal P})
\ee
where the cubic curvature term is
\be
{\cal P} = 12R_\mu{}^\rho{}_\nu{}^\sigma R_\rho{}^\gamma{}_\sigma{}^\delta R_\gamma{}^\mu{}_\delta{}^\nu+R_{\mu\nu}^{\rho\sigma}R_{\rho\sigma}^{\gamma\delta}
R_{\gamma\delta}^{\mu\nu}-12R_{\mu\nu\rho\sigma}R^{\mu\rho}R^{\nu\sigma}
+8R_\mu^\nu R_\nu^\rho R_\rho^\mu\,.
\ee
The covariant equation of motion is
\be
{\cal E}_{ab} \equiv P_{acde}R_b{}^{cde}-\ft12g_{ab}{L}-2\nabla^c\nabla^dP_{acdb} = 0\,,\label{eom}
\ee
where
\bea
P_{abcd} &\equiv& \frac{\partial{L}}{\partial R^{abcd}}\nn\\
&=&\ft12(g_{ac}g_{bd}-g_{ad}g_{bc})+6\lambda
\Big(R_{ad}R_{bc}-R_{ac}R_{bd}+g_{bd}R_a{}^eR_{ce}-g_{ad}R_b{}^eR_{ce}\nn\\
&&-g_{bc}R_a{}^eR_{de}+g_{ac}R_b{}^eR_{de}-g_{bd}R^{ef}R_{aecf}
+g_{bc}R^{ef}R_{aedf}+g_{ad}R^{ef}R_{becf}\nn\\
&&-3R_a{}^e{}_d{}^fR_{becf}-g_{ac}R^{ef}R_{bedf}+3R_a{}^e{}_c{}^fR_{bedf}
+\ft12R_{ab}{}^{ef}R_{cdef}\Big)\,.\label{pabcd}
\eea
The static spherically symmetric BH in ECG is obtained in \cite{Bueno:2016lrh,Hennigar:2016gkm}. The bouncing universe in critical ECG is obtained in \cite{Feng:2017tev}. The slowly rotating BH to leading order in spin for ECG with arbitrary coupling constant of cubic term is obtained in \cite{Adair:2020vso}. 
To study the BZ process in ECG to the second relative order in spin, we need construct the slowly rotating BH in Boyer-Lindquist coordinates, to cubic order in spin and to leading order in coupling constant of cubic terms. Following the scheme in \cite{Yunes:2009hc}, we can expand the metric as
\be
g_{\mu\nu}=g^{(0,0)}_{\mu\nu}+\chi g^{(1,0)}_{\mu\nu}+\chi^2g^{(2,0)}_{\mu\nu}+\chi^3g^{(3,0)}_{\mu\nu}+\zeta g^{(0,1)}_{\mu\nu}+\zeta\chi g^{(1,1)}_{\mu\nu}+\zeta\chi^2g^{(2,1)}_{\mu\nu}+\zeta\chi^3g^{(3,1)}_{\mu\nu}
\ee
where $g^{(0,0)}_{\mu\nu}$ is the Schwarzschild solution, and $g^{(m,n)}_{\mu\nu}$ is a metric perturbation away from the Schwarzschild solution in order ${\cal O}(\zeta^n\chi^m)$. Note that, we have introduced the dimensionless coupling constant $\zeta$ and dimensionless spin $\chi$ by setting
\be
\lambda=M^4\zeta,\quad a=M\chi
\ee
where $M$ is the mass of BH, and $a$ is the spin of BH. After gauge fixing, the metric ansatz for slowly rotating BH take the form
\bea
ds^2&=&-(f+\zeta f^{(0,1)}+\chi^2f^{(2,0)}+\zeta\chi^2f^{(2,1)})dt^2+\frac{dr^2}{f+\zeta g^{(0,1)}+\chi^2g^{(2,0)}+\zeta\chi^2g^{(2,1)}}\nn\\
&&+r^2\big[(1+\chi^2\Theta^{(2,0)}+\zeta\chi^2\Theta^{(2,1)})d\theta^2+(1+\chi^2\Phi^{(2,0)}+\zeta\chi^2\Phi^{(2,1)})\sin^2\theta d\phi^2\big]\nn\\
&&-2(\chi\omega^{(1,0)}+\zeta\chi\omega^{(1,1)}+\chi^3\omega^{(3,0)}+\zeta\chi^3\omega^{(3,1)})\sin^2\theta dtd\phi
\eea
where $\omega^{(1,0)}, f^{(2,0)}, g^{(2,0)}, \Theta^{(2,0)}, \Phi^{(2,0)}$ can be read off by taking the slowly rotating limit of Kerr BH up to the cubic order of spin
\bea
f^{(2,0)}&=&\frac{2M^3}{r^3}\cos^2\theta,\quad g^{(2,0)}=\frac{M^2}{r^2}-\frac{M^2}{r^2}f\cos^2\theta,\quad\Theta^{(2,0)}=\frac{M^2}{r^2}\cos^2\theta\nn\\ \Phi^{(2,0)}&=&\frac{M^2}{r^2}+\frac{2M^3}{r^3}\sin^2\theta,\quad\omega^{(1,0)}=\frac{2M^2}{r},\quad\omega^{(3,0)}=-\frac{2M^4}{r^3}\cos^2\theta
\eea
The $f^{(0,1)}, g^{(0,1)}, \omega^{(1,1)}$ can be obtained by taking small coupling approximation
\be
f^{(0,1)}=g^{(0,1)}=\frac{216 M^6}{r^6}-\frac{368 M^7}{r^7},\quad\omega^{(1,1)}=\frac{368 M^8}{r^7}
\ee
We can decompose the ${\cal O}(\zeta\chi^2)$ and ${\cal O}(\zeta\chi^3)$ perturbations as \cite{Sago:2002fe}
\bea
f^{(2,1)}(r,\theta)&=&H_0^0(r)+H_0^2(r)P_2(\cos\theta)\nn\\
g^{(2,1)}(r,\theta)&=&H_1^0(r)+H_1^2(r)P_2(\cos\theta)\nn\\
\Theta^{(2,1)}(r,\theta)&=&\Phi^{(2,1)}(r,\theta)=H_2^2(r)P_2(\cos\theta)\nn\\
\omega^{(3,1)}(r,\theta)&=&H_3^0(r)+H_3^2P_2(\cos\theta)
\eea
The solution of $H_0^0, H_0^2, H_1^0, H_1^2, H_2^2$ are given in the appendix \ref{appendixA}. The horizon is determined by $g_{tt}g_{\phi\phi}-g_{t\phi}^2=0$, and the solution is
\be
r_H=2M-\chi^2\frac{M}{2}-\zeta M-\zeta\chi^2\frac{11M}{48}+{\cal O}(\zeta^2,\chi^4)
\ee
The angular velocity of the horizon is
\be
\Omega_H\equiv-\frac{g_{tt}}{g_{t\phi}}\Big|_{r=r_H}=\frac{\chi}{4M}+\frac{\chi^3}{16M}+\zeta\chi\frac{35}{32M}+\zeta\chi^3\frac{7}{64M}+{\cal O}(\zeta^2,\chi^5)
\ee
The ergosphere is determined by $g_{tt}=0$
\be
r_{\rm ergo}=2M-\chi^2\frac{M}{2}\cos^2\theta-\zeta M-\zeta\chi^2\frac{11M}{48}(1-15\sin^2\theta)+{\cal O}(\zeta^2,\chi^4)
\ee

\section{Magnetosphere solution in Einsteinian cubic gravity}
We now derive the magnetosphere solution around the slowly rotating BH in ECG by solving the stream equation \eqref{streameq} under the boundary conditions \eqref{bc} and Znajek conditions \eqref{znajek}. First we expand the field variables $\psi, I, \Omega$ as follows \cite{Dong:2021yss}
\bea
\psi&=&\psi^{(0,0)}+\chi^2\psi^{(2,0)}+\zeta\psi^{(0,1)}+\zeta\chi^2\psi^{(2,1)}\nn\\
I&=&\chi i_{(1,0)}(\psi)+\chi^3i_{(3,0)}(\psi)+\zeta\chi i_{(1,1)}(\psi)+\zeta\chi^3i_{(3,1)}(\psi)\nn\\
&=&\chi I^{(1,0)}+\chi^3I^{(3,0)}+\zeta\chi I^{(1,1)}+\zeta\chi^3I^{(3,1)}\nn\\
\Omega&=&\chi \omega_{(1,0)}(\psi)+\chi^3\omega_{(3,0)}(\psi)+\zeta\chi \omega_{(1,1)}(\psi)+\zeta\chi^3\omega_{(3,1)}(\psi)\nn\\
&=&\chi \Omega^{(1,0)}+\chi^3\Omega^{(3,0)}+\zeta\chi \Omega^{(1,1)}+\zeta\chi^3\Omega^{(3,1)}
\eea
It is easy to obtain the expansion coefficients
\bea
I^{(1,0)}&=&i_{(1,0)}(\psi^{(0,0)}),\quad I^{(3,0)}=i_{(3,0)}(\psi^{(0,0)})+i'_{(1,0)}(\psi^{(0,0)})\psi^{(2,0)}\nn\\
I^{(1,1)}&=&i_{(1,1)}(\psi^{(0,0)})+i'_{(1,0)}(\psi^{(0,0)})\psi^{(0,1)},\nn\\
I^{(3,1)}&=&i_{(3,1)}(\psi^{(0,0)})+i'_{(1,0)}(\psi^{(0,0)})\psi^{(2,1)}+i'_{(1,1)}(\psi^{(0,0)})\psi^{(2,0)}\nn\\
&&+i'_{(3,0)}(\psi^{(0,0)})\psi^{(0,1)}+i''_{(1,0)}(\psi^{(0,0)})\psi^{(0,1)}\psi^{(2,0)}
\eea
and similar relations between $\Omega^{(n,m)}$ and $\omega_{(n,m)}$. Then we can compute the field variables $\psi, I, \Omega$ order by order of ${\cal O}(\zeta^n\chi^m)$.

\subsection{leading order in spin}
\begin{center}
\it \underline{Contribution from GR: ${\cal O}(\zeta^0\chi^0)$}
\end{center}
At order ${\cal O}(\zeta^0\chi^0)$, the stream equation is
\be
L\psi^{(0,0)}=0
\ee
where $L$ is a separable differential operator defined by
\be
L=\frac{\partial}{\partial x}\left[(1-\frac{2}{x}\frac{\partial}{\partial x})\right]+\frac{\sin\theta}{x^2}\frac{\partial}{\partial\theta}\left(\frac{1}{\sin\theta}\frac{\partial}{\partial\theta}\right)
\ee
where we have introduced $x \equiv r/M$ as a dimensionless radial coordinate. Imposing the boundary conditions \eqref{bc}, one obtains
\be
\psi^{(0,0)}=\psi_0(1-\cos\theta)
\ee
which is the exact monopole solution. Note that $x_H=2+{\cal O}(\chi^2)$, so we impose the horizon boundary conditions at $x=2$ instead of $x_H$. The Znajek conditions give
\be
\quad I^{(1,0)}=-2\pi\psi_0\Omega^{(1,0)}\sin^2\theta,\quad\Omega^{(1,0)}=\frac{1}{8M}
\ee
These are just the leading order results in GR \cite{Blandford:1977ds,Tanabe:2008wm}.

\begin{center}
\it\underline{Contribution from cubic term: ${\cal O}(\zeta^1\chi^0)$}
\end{center}
At order ${\cal O}(\zeta^1\chi^0)$, the stream equation is the same as the leading equation of GR
\be
L\psi^{(0,1)}=0
\ee
Since the GR solution $\psi^{(0,0)}$ account for all the monopole charge $\psi_0$, we take
\be
\psi^{(0,1)}=0
\ee
The Znajek conditions give
\bea
\quad I^{(1,1)}=-2\pi\psi_0\Omega^{(1,1)}\sin^2\theta,\quad \Omega^{(1,1)}=\frac{35}{64M}
\eea

\subsection{second relative order in spin}
\begin{center}
\it\underline{Contribution from GR: ${\cal O}(\zeta^0\chi^2)$}
\end{center}
At order ${\cal O}(\zeta^0\chi^2)$, the stream equation is
\be
L\psi^{(2,0)}=-\psi_0\frac{x+2}{x^4}\cos\theta\sin^2\theta
\ee
This equation can be solved by separating variables. Imposing the boundary conditions \eqref{bc}, the general solution is
\be
\psi^{(2,0)}=\psi_0\varphi(x)\cos\theta\sin^2\theta
\ee
Substituting it into the stream equation \eqref{streameq}, we obtain the equation for $\varphi(x)$
\be
\frac{d}{dx}\left(\Big(1-\frac2x\Big)\varphi'(x)\right)-\frac{6\varphi(x)}{x^2}+\frac{x+2}{x^4}=0
\ee
The solution of $\varphi(x)$ satisfying the boundary conditions \eqref{bc} is
\bea
\varphi(x)&=&\frac18x^2(2x-3)\left[\text{Li}_2\Big(\frac2x\Big)+\ln\Big(\frac2x\Big)\ln\Big(1-\frac2x\Big)\right]\nn\\
&&+\frac{1}{12}(6x^2-3x-1)\ln\Big(\frac2x\Big)-\frac16x^2(x-1)+\frac{11}{72}+\frac{1}{3x}
\eea
The Znajek conditions give
\bea
I^{(3,0)}&=&-2\pi\psi_0\sin^2\theta\left[\Omega^{(3,0)}+2\Omega^{(1,0)}\varphi(x)\right]\nn\\
\Omega^{(3,0)}&=&\frac{1}{32M}+\frac{67-6\pi^2}{1152M}\sin^2\theta
\eea
These are just the second relative order results in GR \cite{Tanabe:2008wm,Armas:2020mio}.

\begin{center}
\it\underline{Contribution from cubic term: ${\cal O}(\zeta^1\chi^2)$}
\end{center}
At order ${\cal O}(\zeta^1\chi^2)$, the stream equation is
\be
L\psi^{(2,1)}=-\psi_0s(x)\cos\theta\sin^2\theta \label{sxeq}
\ee
where $s(x)$ is given by \eqref{sx} in appendix \ref{appendixB}. Similarly the general solution of $\psi^{(2,1)}$ takes the form
\be
\psi^{(2,1)}=\psi_0\varphi_\zeta(x)\cos\theta\sin^2\theta
\ee
and $\varphi_\zeta(x)$ is determined by the equation
\be
\frac{d}{dx}\left(\Big(1-\frac2x\Big)\varphi_\zeta'(x)\right)-\frac{6\varphi_\zeta(x)}{x^2}+s(x)=0\label{varphieq}
\ee
whose solution satisfying the boundary conditions \eqref{bc} is given by \eqref{varphix} in appendix \ref{appendixB}. The Znajek conditions give
\bea
I^{(3,1)}&=&-2\pi\psi_0\sin^2\theta\Big[\Omega^{(3,1)}+2\Omega^{(1,0)}\varphi_\zeta\cos^2\theta+2\Omega^{(1,1)}\varphi\cos^2\theta\Big]\nn\\
\Omega^{(3,1)}&=&\frac{7}{128M}+\frac{120543743-9101400 \pi ^2 }{129024000 M}\sin^2\theta
\eea

\section{The power of energy extraction}
Up to second relative order in spin, the power of energy extraction can be obtained by \eqref{BZpower}
\be
P=\left[\frac{\pi}{24}+\frac{35\pi}{96}\zeta\right]\frac{\psi_0^2\chi^2}{M^2}+\left[\frac{\pi\left(56-3\pi^2\right)}{1080}+\frac{\pi\left(61414581-4013800\pi^2\right)}{80640000}\zeta\right]\frac{\psi_0^2\chi^4}{M^2}
\ee
In order to discuss our result, we arrange the horizon angular velocity in the same relative order
\bea
\Omega_H=\left[\frac{1}{4}+\frac{35}{32}\zeta\right]\frac{\chi}{M}+\left[\frac{1}{16}+\frac{7}{64}\zeta\right]\frac{\chi^3}{M}+{\cal O}(\zeta^2,\chi^5)
\eea
and the angular velocity of the magnetic field lines
\be
\Omega=\left[\frac{1}{8}+\frac{35}{64}\zeta\right]\frac{\chi}{M}+\left[\frac{1}{32}+\frac{7}{128}\zeta+\left(\frac{67-6\pi^2}{1152}+\frac{120543743-9101400 \pi ^2 }{129024000}\zeta\right)\sin^2\theta\right]\frac{\chi^3}{M}
\ee
The leading BZ power correction with respect to GR is
\be
\frac{P_{\rm ECG}-P_{\rm GR}}{P_{\rm GR}}\approx27.5\zeta
\ee
For given mass of BH $M$ and magnetic flux $\psi_0$, our result shows that the BZ power is dependent on two parameters: the coupling constant $\zeta$ and the spin of BH $\chi$. So if we have independent and high quality measurements of the jet power and the spin of BH, we can distinguish GR from other theories of gravity by fitting data in principle. However even within
GR, a clear observational signature of the BZ mechanism is still missing. Analytical calculation and numerical stimulations show that for general rotating BH the maximum rate of energy extraction is achieved when $\Omega=\ft12\Omega_H$, and takes the form
\be
P=k\Omega_H^2\Phi_{\rm BH}^2
\ee
where $k=1/6\pi$ for a split monopole field profile and $k = 0.044$ for a paraboloidal profile, and $\Phi_{\rm BH}$ is the total magnetic flux through the BH horizon. Numerical stimulations show that this result is accurate even for large spins up to $\chi\le0.95$ for Kerr black hole \cite{Tchekhovskoy:2009ba,Tchekhovskoy:2011zx}. It is easy to check that this general character of BZ process still holds at leading order in spin for ECG, i.e.
\bea
\Omega&=&\fft12\Omega_H+{\cal O}(\zeta^2,\chi^3)\nn\\
P&=&\frac{1}{6\pi}\Omega_H^2\Phi_{\rm BH}^2+{\cal O}(\zeta^2,\chi^4)
\eea
where $\Phi_{\rm BH}=2\pi\psi_0$. Since $P$ is a function which only depends on $\Omega_H$ at leading order in spin, the two parameters $\zeta$ and $\chi$ are degenerate to this order. This implies that we will not be able to determine both the coupling constant $\zeta$ and the spin of BH $\chi$ even if both the BZ power $P$ and the horizon angular velocity $\Omega_H$ are measured. However this degeneracy breaks when higher orders in spin are considered. We can see this by rewriting the BZ power as
\be
P=\frac{1}{6\pi}\frac{\Phi^2}{M^2}[(M\Omega_H)^2+(\alpha+\beta\zeta)(M\Omega_H)^4]+{\cal O}(\zeta^2,\chi^6)
\ee
where
\bea
\alpha&=&\frac{8(67-6\pi^2)}{45}\approx1.38\nn\\
\beta&=&\frac{22053743}{630000}-\frac{67 \pi ^2}{150}\approx30.60
\eea
When $\zeta=0$, we recover the GR result which is accurate for large spins up to $\chi\le0.99$ Kerr black hole \cite{Tchekhovskoy:2009ba,Tchekhovskoy:2011zx}. These similar results were obtained for quadratic gravities in \cite{Dong:2021yss}.


\section{Conclusion}
The BZ process can be viewed as a magnetic version of Penrose process, whose primary mechanism is frame dragging due to the rotation of BH \cite{Dadhich:2018gmh,Tursunov:2020juz} (One can see detail discussions on the motion of charged particle around a magnetized Kerr BH in \cite{Hou:2023hto,Zhang:2023cuw}). In general, a magnetic version of Penrose process depend both on the spacetime geometry and the magnetic field configuration. Since the accretion disk is typically a low-density plasma, with the energy of magnetic field overwhelming the energy of accretion disk, we can neglect the energy exchange between magnetic field and accretion disk. Thus in the BZ process, we can determine the magnetosphere around the BH based on solving the force-free electrodynamics under reasonable boundary conditions. Theoretical analysis and numerical stimulation of this model in Kerr BH indicate that the power of outflowing jet depends on both the angular velocity of BH horizon $\Omega_{\rm H}$ and the amount of magnetic flux threading the BH horizon $\Phi_{\rm BH}$, with jet power $\propto\Omega_{\rm H}^2\Phi_{\rm BH}^2$.The Event Horizon Telescope (EHT) observation of M87$^\ast$, especially the polarization data favors this model strongly. \cite{EventHorizonTelescope:2019pgp,EventHorizonTelescope:2021srq}.

In this paper, we study the BZ process in the framework of ECG. As expected, the BZ power is modified by the coupling constant of higher curvature terms $\zeta$. However testing this result with jet power is extremely challenging, even though the BZ mechanism is well understood, since we need so many data to be measured independently, especially the BH spin $\chi$. In fact our result show that the BH spin $\chi$ and coupling constant $\zeta$ are degenerate in the BZ power at leading order in spin. The degeneracies between the astrophysical or theory-dependent parameter and BH spin are common for slow rotating BH. However, in ECG and quadratic gravities, this degeneracy breaks down at higher orders in spin. To summarize, the feature of BZ power in higher derivative gravities deviate from the one in GR for BH with large spin. Consequently, we expect that it is possible to distinguish GR from higher derivative gravities using jet power with rapidly rotating BH.

\section*{Acknowledgments}
X.H.F. is supported by NSFC (National Natural Science Foundation of China) Grant No. 11905157 and No. 11935009.
\appendix

\section{The metric functions at orders ${\cal O}(\zeta\chi^2)$ and ${\cal O}(\zeta\chi^3)$}\label{appendixA}
The metric functions $H_0^0, H_0^2, H_1^0, H_1^2, H_2^2, H_3^0, H_3^2$ can be obtained by solving the equation of motion \eqref{eom} up to the order ${\cal O}(\zeta\chi^3)$, and are given by
\begin{footnotesize}
\bea
H_0^0&=&-\frac{368 M^{10}}{3 r^{10}}+\frac{7592 M^9}{3 r^9}-\frac{1512 M^8}{r^8},\quad H_1^0=\frac{1840 M^{10}}{r^{10}}+\frac{7672 M^9}{3 r^9}-\frac{1728 M^8}{r^8}\nn\\
H_0^2&=&-\frac{2208 M^{11}}{r^{11}}-\frac{48 M^{10}}{r^{10}}+\frac{16480 M^9}{3 r^9}-\frac{2992 M^8}{r^8}+\frac{160 M^7}{7 r^7}+\frac{120 M^6}{7 r^6}+\frac{96 M^5}{7 r^5}+\frac{12 M^4}{r^4}+\frac{12 M^3}{r^3}\nn\\
H_1^2&=&\frac{37536 M^{11}}{r^{11}}-\frac{19152 M^{10}}{r^{10}}+\frac{3232 M^9}{3 r^9}-\frac{1264 M^8}{r^8}+\frac{160 M^7}{7 r^7}+\frac{120 M^6}{7 r^6}+\frac{96 M^5}{7 r^5}+\frac{12 M^4}{r^4}+\frac{12 M^3}{r^3}\nn\\
H_2^2&=&-\frac{2208 M^{10}}{r^{10}}-\frac{1616 M^9}{3 r^9}-\frac{1600 M^7}{7 r^7}-\frac{120 M^6}{r^6}-\frac{432 M^5}{7 r^5}-\frac{30 M^4}{r^4}-\frac{12 M^3}{r^3}\nn\\
H_3^0&=&-\frac{368 M^{11}}{3 r^{10}}-\frac{8024 M^{10}}{3 r^9}+\frac{352 M^9}{7 r^8}+\frac{160 M^8}{7 r^7}+\frac{60 M^7}{7 r^6}+\frac{9 M^6}{7 r^5}-\frac{31 M^5}{14 r^4}-\frac{23 M^4}{7 r^3}\nn\\
H_3^2&=&\frac{2208 M^{12}}{r^{11}}-\frac{1328 M^{11}}{3 r^{10}}-\frac{17344 M^{10}}{3 r^9}+\frac{480 M^9}{7 r^8}+\frac{240 M^8}{7 r^7}+\frac{12 M^7}{r^6}-\frac{15 M^6}{7 r^5}-\frac{155 M^5}{14 r^4}-\frac{115 M^4}{7 r^3}\nn\\
\eea
\end{footnotesize}

\section{Solution of the stream equation at order ${\cal O}(\zeta\chi^2)$}\label{appendixB}
The function $s(x)$ in the stream equation at order ${\cal O}(\zeta\chi^2)$ \eqref{sxeq} is
\begin{footnotesize}
\bea
s(x)&=&\frac{1}{168(x-2)^2}\left(-\frac{24482304}{x^{12}}+\frac{23256576}{x^{11}}-\frac{3459456}{x^{10}}-\frac{1484032}{x^9}+\frac{1105024}{x^8}-\frac{3519424}{x^7}\right.\nn\\
&&\qquad\qquad\qquad\left.+\frac{3667536}{x^6}-\frac{1492704}{x^5}+\frac{226800}{x^4}-\frac{6048}{x^3}-\frac{1470}{x^2}+\frac{735}{x}\right)\nn\\
&&-\frac{4}{(x-2)^2}\left(\frac{644}{x^8}-\frac{3728}{x^7}+\frac{4545}{x^6}-\frac{2067}{x^5}+\frac{324}{x^4}\right)\ln\Big(\frac2x\Big)\nn\\
&&-6\left(\frac{276}{x^7}-\frac{365}{x^6}+\frac{108}{x^5}\right)\left[\ln\Big(\frac2x\Big)\ln\Big(1-\frac2x\Big)+\text{Li}_2\Big(\frac2x\Big)\right]\label{sx}
\eea
\end{footnotesize}
The solution $\varphi_\zeta(x)$ of equation \eqref{varphieq} is
\begin{footnotesize}
\bea
\varphi_\zeta(x)&=&\frac{x^4}{2016000 (x - 2)}\left(\frac{741888000}{x^{13}}+\frac{64512000}{x^{12}}-\frac{57984000}{x^{11}}-\frac{12784000}{x^{10}}-\frac{97814080}{x^9}+\frac{449554800}{x^8}\right.\nn\\
&&\qquad\qquad\qquad\qquad-\frac{339643816}{x^7}+\frac{78226442}{x^6}-\frac{13759137}{x^5}-\frac{1754200}{x^4}-\frac{15094450}{x^3}+\frac{26724600}{x^2}\nn\\
&&\qquad\qquad\qquad\qquad\left.-\frac{8908200}{x}\right)+\frac{x^4}{4800(x-2)}\left(\frac{88320}{x^9}-\frac{754800}{x^8}+\frac{731016}{x^7}-\frac{182532}{x^6}+\frac{2142}{x^5}\right.\nn\\
&&\qquad\qquad\qquad\qquad\qquad\qquad\qquad\qquad\qquad\qquad\quad\left.+\frac{7070}{x^4}+\frac{17675}{x^3}-\frac{53025}{x^2}+\frac{21210}{x}\right)\ln\Big(\frac2x\Big)\nn\\
&&-\frac{x^4}{640}\left(\frac{22080}{x^8}-\frac{36416}{x^7}+\frac{12264}{x^6}+\frac{2121}{x^2}-\frac{1414}{x}\right)\left[\ln\Big(\frac2x\Big)\ln\Big(1-\frac2x\Big)+\text{Li}_2\Big(\frac2x\Big)\right]\label{varphix}
\eea
\end{footnotesize}


\begin{thebibliography}{10}

\bibitem{Blandford:1977ds}
R.~D.~Blandford and R.~L.~Znajek,
``Electromagnetic extractions of energy from Kerr black holes,''
Mon. Not. Roy. Astron. Soc. \textbf{179}, 433-456 (1977)

\bibitem{Beskin:2000qe}
V.~S.~Beskin and I.~V.~Kuznetsova,
``On the blandford - znajek mechanism of the energy loss of a rotating black hole,''
Nuovo Cim. B \textbf{115}, 795 (2000)
[arXiv:astro-ph/0004021 [astro-ph]].

\bibitem{McKinney:2004ka}
J.~C.~McKinney and C.~F.~Gammie,
``A Measurement of the electromagnetic luminosity of a Kerr black hole,''
Astrophys. J. \textbf{611}, 977-995 (2004)
[arXiv:astro-ph/0404512 [astro-ph]].

\bibitem{Tanabe:2008wm}
K.~Tanabe and S.~Nagataki,
``Extended monopole solution of the Blandford-Znajek mechanism: Higher order terms for a Kerr parameter,''
Phys. Rev. D \textbf{78}, 024004 (2008)
[arXiv:0802.0908 [astro-ph]].

\bibitem{Tchekhovskoy:2009ba}
A.~Tchekhovskoy, R.~Narayan and J.~C.~McKinney,
``Black Hole Spin and the Radio Loud/Quiet Dichotomy of Active Galactic Nuclei,''
Astrophys. J. \textbf{711}, 50-63 (2010)
[arXiv:0911.2228 [astro-ph.HE]].

\bibitem{Pan:2015haa}
Z.~Pan and C.~Yu,
``Fourth-order split monopole perturbation solutions to the Blandford-Znajek mechanism,''
Phys. Rev. D \textbf{91}, no.6, 064067 (2015)
[arXiv:1503.05248 [astro-ph.HE]].

\bibitem{Pan:2015iaa}
Z.~Pan and C.~Yu,
``Analytic Properties of Force-free Jets in the Kerr Spacetime\textemdash{}I,''
Astrophys. J. \textbf{812}, no.1, 57 (2015)
[arXiv:1504.04864 [astro-ph.HE]].

\bibitem{Grignani:2018ntq}
G.~Grignani, T.~Harmark and M.~Orselli,
``Existence of the Blandford-Znajek monopole for a slowly rotating Kerr black hole,''
Phys. Rev. D \textbf{98}, no.8, 084056 (2018)
[arXiv:1804.05846 [gr-qc]].

\bibitem{Grignani:2019dqc}
G.~Grignani, T.~Harmark and M.~Orselli,
``Force-free electrodynamics near rotation axis of a Kerr black hole,''
Class. Quant. Grav. \textbf{37}, no.8, 085012 (2020)
[arXiv:1908.07227 [gr-qc]].

\bibitem{Armas:2020mio}
J.~Armas, Y.~Cai, G.~Comp\`ere, D.~Garfinkle and S.~E.~Gralla,
``Consistent Blandford-Znajek Expansion,''
JCAP \textbf{04}, 009 (2020)
[arXiv:2002.01972 [astro-ph.HE]].

\bibitem{Camilloni:2022kmx}
F.~Camilloni, O.~J.~C.~Dias, G.~Grignani, T.~Harmark, R.~Oliveri, M.~Orselli, A.~Placidi and J.~E.~Santos,
``Blandford-Znajek monopole expansion revisited: novel non-analytic contributions to the power emission,''
JCAP \textbf{07}, no.07, 032 (2022)
[arXiv:2201.11068 [gr-qc]].

\bibitem{Komissarov:2001}
S.~S.~Komissarov,
''Direct numerical simulations of the Blandford-Znajek effect,''
Mon. Not. R. Astron. Soc. \textbf{326}, L41-L44 (2001)

\bibitem{Komissarov:2004qu}
S.~S.~Komissarov,
``General relativistic MHD simulations of monopole magnetospheres of black holes,''
Mon. Not. Roy. Astron. Soc. \textbf{350}, 1431 (2004)
[arXiv:astro-ph/0402430 [astro-ph]].

\bibitem{Tchekhovskoy:2011zx}
A.~Tchekhovskoy, R.~Narayan and J.~C.~McKinney,
``Efficient Generation of Jets from Magnetically Arrested Accretion on a Rapidly Spinning Black Hole,''
Mon. Not. Roy. Astron. Soc. \textbf{418}, L79-L83 (2011)
[arXiv:1108.0412 [astro-ph.HE]].

\bibitem{Nathanail:2014aua}
A.~Nathanail and I.~Contopoulos,
``Black Hole Magnetospheres,''
Astrophys. J. \textbf{788}, no.2, 186 (2014)
[arXiv:1404.0549 [astro-ph.HE]].

\bibitem{Bambi:2012zg}
C.~Bambi,
``Attempt to find a correlation between the spin of stellar-mass black hole candidates and the power of steady jets: relaxing the Kerr black hole hypothesis,''
Phys. Rev. D \textbf{86}, 123013 (2012)
[arXiv:1204.6395 [gr-qc]].

\bibitem{Pei:2016kka}
G.~Pei, S.~Nampalliwar, C.~Bambi and M.~J.~Middleton,
``Blandford-Znajek mechanism in black holes in alternative theories of gravity,''
Eur. Phys. J. C \textbf{76}, no.10, 534 (2016)
[arXiv:1606.04643 [gr-qc]].

\bibitem{Konoplya:2021qll}
R.~A.~Konoplya, J.~Kunz and A.~Zhidenko,
``Blandford-Znajek mechanism in the general stationary axially-symmetric black-hole spacetime,''
JCAP \textbf{12}, no.12, 002 (2021)
[arXiv:2102.10649 [gr-qc]].

\bibitem{Banerjee:2020ubc}
I.~Banerjee, B.~Mandal and S.~SenGupta,
``Signatures of Einstein-Maxwell dilaton-axion gravity from the observed jet power and the radiative efficiency,''
Phys. Rev. D \textbf{103}, no.4, 044046 (2021)
[arXiv:2007.03947 [gr-qc]].

\bibitem{Dong:2021yss}
J.~Dong, N.~Pati\~no, Y.~Xie, A.~C\'ardenas-Avenda\~no, C.~F.~Gammie and N.~Yunes,
``Blandford-Znajek process in quadratic gravity,''
Phys. Rev. D \textbf{105}, no.4, 044008 (2022)
[arXiv:2111.08758 [gr-qc]].

\bibitem{Horowitz:2023xyl}
G.~T.~Horowitz, M.~Kolanowski, G.~N.~Remmen and J.~E.~Santos,
``Extremal Kerr black holes as amplifiers of new physics,''
[arXiv:2303.07358 [hep-th]].

\bibitem{Bueno:2016xff}
P.~Bueno and P.~A.~Cano,
``Einsteinian cubic gravity,''
Phys. Rev. D \textbf{94}, no.10, 104005 (2016)
[arXiv:1607.06463 [hep-th]].

\bibitem{Gralla:2014yja}
S.~E.~Gralla and T.~Jacobson,
``Spacetime approach to force-free magnetospheres,''
Mon. Not. Roy. Astron. Soc. \textbf{445}, no.3, 2500-2534 (2014)
[arXiv:1401.6159 [astro-ph.HE]].

\bibitem{Znajek(1977)}
R.~L.~Znajek,
''Black hole electrodynamics and the Carter tetrad,''
Monthly Notices of the Royal Astronomical Society \textbf{179}, 457-472 (1977).

\bibitem{Bueno:2016lrh}
P.~Bueno and P.~A.~Cano,
``Four-dimensional black holes in Einsteinian cubic gravity,''
Phys. Rev. D \textbf{94}, no.12, 124051 (2016)
[arXiv:1610.08019 [hep-th]].

\bibitem{Hennigar:2016gkm}
R.~A.~Hennigar and R.~B.~Mann,
``Black holes in Einsteinian cubic gravity,''
Phys. Rev. D \textbf{95}, no.6, 064055 (2017)
[arXiv:1610.06675 [hep-th]].

\bibitem{Feng:2017tev}
X.~H.~Feng, H.~Huang, Z.~F.~Mai and H.~Lu,
``Bounce Universe and Black Holes from Critical Einsteinian Cubic Gravity,''
Phys. Rev. D \textbf{96}, no.10, 104034 (2017)
[arXiv:1707.06308 [hep-th]].

\bibitem{Adair:2020vso}
C.~Adair, P.~Bueno, P.~A.~Cano, R.~A.~Hennigar and R.~B.~Mann,
``Slowly rotating black holes in Einsteinian cubic gravity,''
Phys. Rev. D \textbf{102}, no.8, 084001 (2020)
[arXiv:2004.09598 [gr-qc]].

\bibitem{Yunes:2009hc}
N.~Yunes and F.~Pretorius,
``Dynamical Chern-Simons Modified Gravity. I. Spinning Black Holes in the Slow-Rotation Approximation,''
Phys. Rev. D \textbf{79}, 084043 (2009)
[arXiv:0902.4669 [gr-qc]].

\bibitem{Sago:2002fe}
N.~Sago, H.~Nakano and M.~Sasaki,
``Gauge problem in the gravitational selfforce. 1. Harmonic gauge approach in the Schwarzschild background,''
Phys. Rev. D \textbf{67}, 104017 (2003)
[arXiv:gr-qc/0208060 [gr-qc]].

\bibitem{Dadhich:2018gmh}
N.~Dadhich, A.~Tursunov, B.~Ahmedov and Z.~Stuchl\'\i{}k,
``The distinguishing signature of Magnetic Penrose Process,''
Mon. Not. Roy. Astron. Soc. \textbf{478}, no.1, L89-L94 (2018)
[arXiv:1804.09679 [astro-ph.HE]].

\bibitem{Tursunov:2020juz}
A.~Tursunov, Z.~Stuchl\'\i{}k, M.~Kolo\v{s}, N.~Dadhich and B.~Ahmedov,
``Supermassive Black Holes as Possible Sources of Ultrahigh-energy Cosmic Rays,''
Astrophys. J. \textbf{895}, no.1, 14 (2020)
[arXiv:2004.07907 [astro-ph.HE]].

\bibitem{Hou:2023hto}
Y.~Hou, Z.~Zhang, M.~Guo and B.~Chen,
``Electromagnetic effects on charged particles in NHEK,''
[arXiv:2301.08467 [gr-qc]].

\bibitem{Zhang:2023cuw}
Z.~Zhang, Y.~Hou, Z.~Hu, M.~Guo and B.~Chen,
``Polarized images of charged particles in vortical motions around a magnetized Kerr black hole,''
[arXiv:2304.03642 [gr-qc]].

\bibitem{EventHorizonTelescope:2019pgp}
K.~Akiyama \textit{et al.} [Event Horizon Telescope],
``First M87 Event Horizon Telescope Results. V. Physical Origin of the Asymmetric Ring,''
Astrophys. J. Lett. \textbf{875}, no.1, L5 (2019)
[arXiv:1906.11242 [astro-ph.GA]].

\bibitem{EventHorizonTelescope:2021srq}
K.~Akiyama \textit{et al.} [Event Horizon Telescope],
``First M87 Event Horizon Telescope Results. VIII. Magnetic Field Structure near The Event Horizon,''
Astrophys. J. Lett. \textbf{910}, no.1, L13 (2021)
[arXiv:2105.01173 [astro-ph.HE]].
\end{thebibliography}
\end{document}